\documentclass[twocolumn,twoside,slac_two]{revtex4}
\pdfoutput=1
\usepackage{graphicx}
\usepackage{fancyhdr}
\begin{document}

\title{First minimum bias physics results at LHCb}

%

\author{Christian Linn, \it{on behalf of LHCb collaboration}}
\affiliation{Physikalisches Institut der Universit\" at Heidelberg, Germany}
%

\begin{abstract}

The goal of the LHCb experiment is the indirect search for New Physics through precision measurements of B-decays.
A short description of the detector and its performance after the first data taking in 2009 and 2010 will be presented.
In addition first preliminary results of a $\rm K_{s}$ differential cross section measurement at a center of mass energy of $\sqrt{s} = 900\; \rm GeV$ 
and a measurement of the $\frac{\bar{\lambda}}{\lambda}$ production ratio at $\sqrt{s} = 900\; \rm GeV$ and $\sqrt{s} = 7\; \rm TeV$ will be discussed.

\end{abstract}

\maketitle

\thispagestyle{fancy}


\section{Introduction}
LHCb is a dedicated flavour-physics experiment to search for New Physics beyond the Standard Model through precision measurements of b hadron decays and 
CP violation. The large $\rm b\bar{b}$ production cross section at the LHC leads to $\sim 10^{12} \; \rm b\bar{b}$ pairs per nominal operational 
($10^7 \; \rm s$) year at a luminosity of $2 \cdot 10^{32} \; \rm cm^{-2}s^{-1}$. 
The dominant production mechanism is through gluon-gluon fusion with asymmetric momenta of the incoming gluons. 
As a consequence the $\rm b\bar{b}$ pair is strongly boosted either in forward or backward direction with respect to the $pp$ center of mass system. 
Similar considerations hold for the production of charm mesons, with which many interesting measurements are also foreseen.
Therefor the LHCb detector is designed as single arm forward spectrometer covering a pseudorapidity range $1.9 < \eta < 4.9$. Beside of the 
good acceptance for detecting b hadrons, this unique angular coverage gives the possibility for a wide field of forward minimum bias physics.\\
Hence, the first data collected in 2009 and 2010 at center of mass energies $\sqrt{s} = 900\; \rm GeV$ and $\sqrt{s} = 7\; \rm TeV$ were not only used to 
calibrate and understand the detector but also to obtain interesting forward physics results. At LHC especially strangeness production is an excellent 
test-field of hadronization and fragmentation models as there are no valence strange quarks in the initial state. 
So far there are several models available which disagree mainly in forward phase space regions. This makes the LHCb results to be a good probe of 
these models. As first LHCb measurements the differential $\rm K_{s}$ production cross section and the 
$\frac{\bar{\lambda}}{\lambda}$ production ratio will be presented here.     

\section{Detector description and performance}
\begin{figure}
\includegraphics[width=90mm]{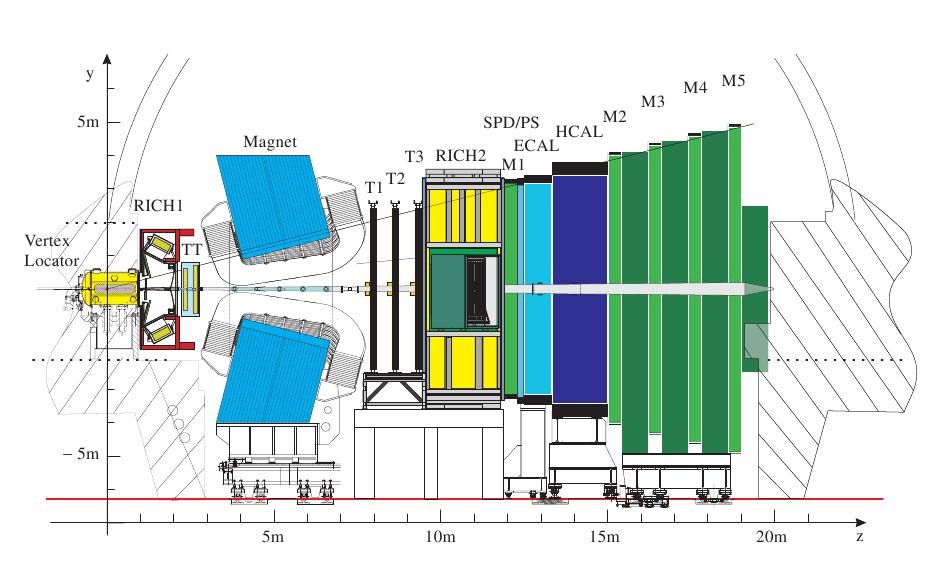}
\caption{Side view of the LHCb detector showing the vertex locator (VELO), the Cherenkov detectors (RICH1, RICH2), the 
dipole magnet and the tracking stations (TT, T1-T3), the Scintillating Pad Detector (SPD), Preshower (PS), electromagnetic (ECAL) and hadronic (HCAL) calorimeters, 
and the five muon stations(M1-M5).}
\label{lhcb_layout}
\end{figure}

The aim of the LHCb detector to measure precisely the b hadrons and their decay products leads to specific detector requirements. An excellent vertex 
and proper time resolution is necessary to measure the fast oscillation of the $\rm B_{s}$ mesons. An efficient particle identification, especially 
for $\rm K/\pi$ separation, and an excellent momentum resolution for the precise invariant mass reconstruction is needed to reject background.\\
The detector consists of a vertex locator (VELO), two Cherenkov detectors (RICH1-2) for particle identification, 
a tracking system with two stations before (TT) and three stations after (T1-3) a warm dipole magnet, electromagnetic (ECAL) and hadronic (HCAL) 
calorimeters and five muon stations (M1-5). A schematic layout of the detector is shown in Fig. \ref{lhcb_layout}.
In the following a few detector components and their performance with the first LHC data are described in more detail. A complete description of the 
detector can be found in \cite{detector}.

\subsection{VELO and tracking system}
The vertex locator is a silicon strip detector built of 21 stations positioned around the interaction region. The stations are made of half disks which 
measure the radial and azimuthal coordinates. The sensitive area of the sensors start at $8 \;\rm mm$ from the beam axis. This small distance allows an 
excellent vertex resolution. Already in the first data a vertex resolution in the plane perpendicular to the beam axis of $\sim 15\;\mu m$ was achieved. 
To avoid damage when setting up the LHC beams the half disks can be retracted.\\ 
The spectrometer consists of a warm dipole magnet with an integrated magnetic field of $\sim 4 \;\rm Tm$. Five tracking stations are used 
for a precise momentum measurement. The two stations before the magnet (TT) consist of four layers of silicon strip detectors. The three stations after the 
magnetic field are divided into two parts: the area close to the beam pipe (IT) covers most of the particle flux and is also made out of 
silicon strip detectors, whereas the outer part (OT) is constructed from drift straw tubes. The hit resolution of the silicon strip detectors determined with 
the first collision data is $\sim 60 \;\rm \mu m$ for the TT and $\sim 55 \;\rm \mu m$ for the IT. The straw tube detector has a hit resolution of $\sim 270 \;\rm \mu m$, 
which is already close to the nominal value.\\
The expected momentum resolution of the tracking system, predicted from Monte Carlo, starts from $\frac{\delta p}{p} \sim 0.35\%$ for low momentum tracks and degrades for high 
momentum tracks up to $\frac{\delta p}{p} \sim 0.55\%$. 
 
\subsection{Particle identification} 
Ring imaging Cherenkov detectors (RICH) are used for particle identification. An efficient $\rm K/\pi$ separation is important for the 
necessary background rejection. The first RICH is located directly after the vertex locator. It uses $\rm C_{4}F_{10}$ gas and aerogel 
as radiators and covers a momentum range from $2 - 60\; \rm GeV$. The second RICH using $\rm C F_{4}$ as gas radiator is positioned after the 
tracking stations and extends the momentum coverage up to $\sim 100\; \rm GeV$. The average efficiency for kaon identification is 
$\epsilon (K \rightarrow K) \sim 95\%$ with a corresponding pion misidentification rate of $\epsilon (\pi \rightarrow K) \sim 5\%$.
Electrons, photons and hadrons are identified with the calorimeter system, which is also used to measure their energies and positions. It is
 positioned after the RICH2 and consists of a scintillating pad detector, a preshower detector, the electromagnetic and hadronic calorimeter. The muon identification is done with the five muon stations, four of them located after the calorimeters. 
The muon identification efficiency $\epsilon (\mu) \sim 97\%$ was estimated from data and is in very good agreement with the Monte Carlo predictions.

\section{$\rm K_{s}$ cross section measurement}
The measurement of the differential $\rm K_{s}$ production cross section was done with the first data delivered by the LHC. In total 13 runs, taken in
December 2009 with an integrated luminosity of $L = 6.8\pm1.0\; \mu b^{-1}$ at a center of mass energy of $\sqrt{s} = 900\; \rm GeV$ were
 used for this analysis. All detector components were in operation and the magnetic field was at nominal strength. However, due to the low beam rigidity at lower energy 
the vertex detector halves were retracted $15\; \rm mm$ from their nominal position.   
\subsection{Analysis strategy}
$K_{s}$ mesons were reconstructed in the $\pi^+\pi^-$ mode using only events triggered by the calorimeters. Contributions from secondary interactions 
in the detector material or from the decay of long-lived particles were suppressed by requiring the $\rm K_{s}$ candidates to point back to the $pp$ collision point. 
The analysis was performed in bins of transverse momentum $p_{\rm T}$ and rapidity $y = \frac{1}{2} \ln \frac{E+p_z}{E-p_z}$, calculated in the rest frame of the $pp$ collision. The partial $K_{s}$ 
cross section was calculated using $\sigma_{i,j} = \frac{N_{i,j}}{L \epsilon_{i,j}^{\rm trig} \epsilon_{i,j}^{\rm reco}}$ where $N_{i,j}$ is the 
observed signal yield in the respective $(p_{\rm T}, y)$ bin and $\epsilon_{i,j}^{\rm trig}$ and $\epsilon_{i,j}^{\rm reco}$ are the corresponding 
trigger and reconstruction efficiencies. The efficiencies were calculated from a fully simulated Monte Carlo sample of single $pp$ collisions and checked with data. 
Details are described in \cite{Kspaper}. No separation of $\rm K_{s}$ produced in diffractive and non-diffractive events was attempted. 

\subsection{Luminosity measurement}
The integrated luminosity was determined using measurements of beam currents and the size of the colliding beam bunches. The luminosity produced by one pair of 
colliding bunches can be expressed as
\begin{equation}
 L = f \sum \frac{n_{1} n_{2}}{4 \pi \sigma^{x} \sigma^{y}}
\end{equation}
where $n_{1}$,$n_{2}$ are the numbers of protons in bunch $1$ and $2$, $f = 11.245 \rm kHz$ is the LHC revolution frequency and 
$\sigma^{x}$, $\sigma^{y}$ are the bunch sizes in the transverse $(x,y)$ plane. The beam sizes were reconstructed using tracks 
produced in beam-beam and beam-gas collisions in the VELO and the bunch currents are measured by the LHC machine team. This procedure has a systematic uncertainty $\sim 15\%$ which is dominated by the measurement of the beam currents delivered from the LHC machine. 

\subsection{Event selection}
\begin{figure}
\includegraphics[width=75mm]{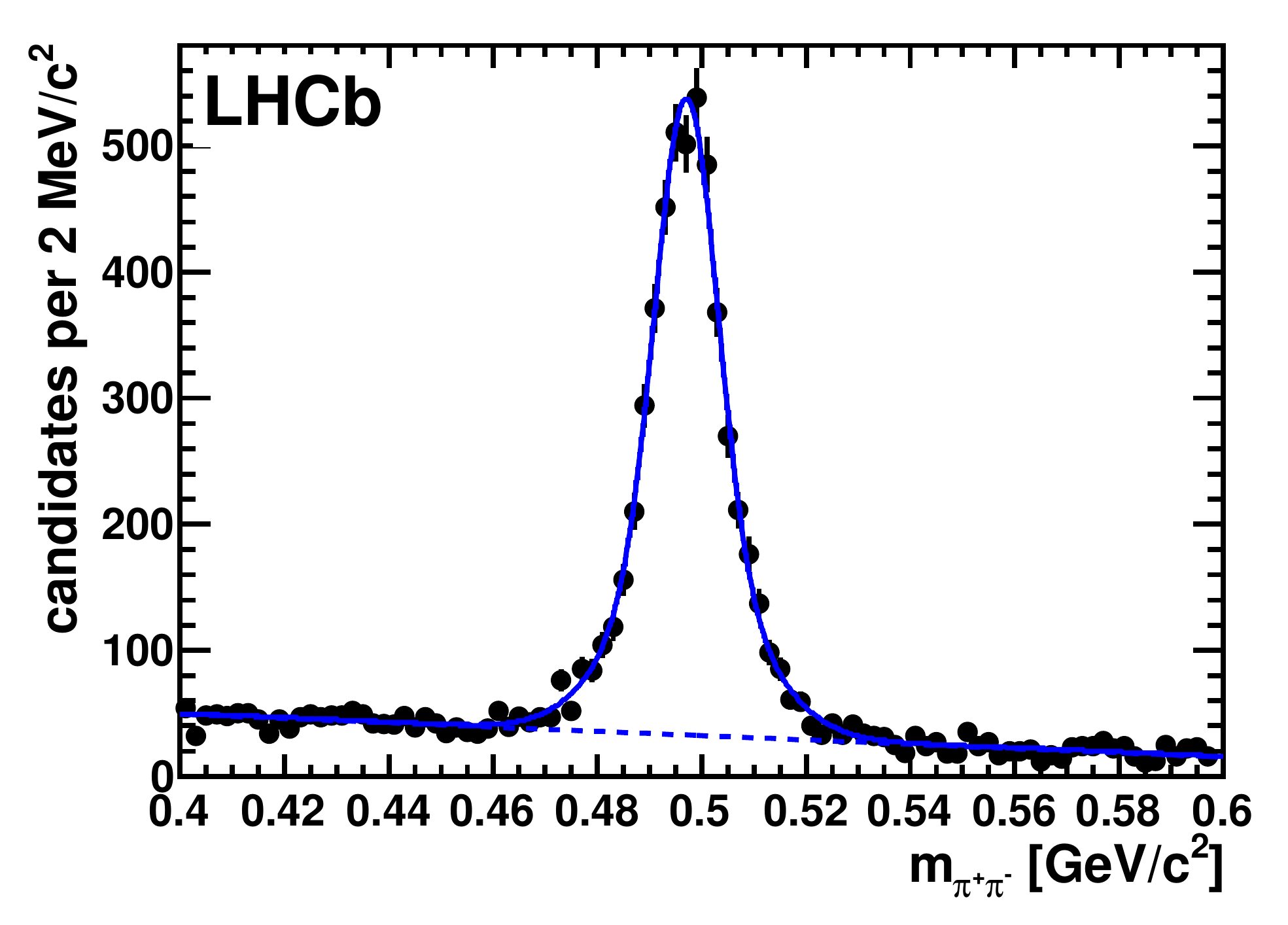}
\caption{Mass distribution of all selected $K_s$ candidates in the downstream-track analysis with the overlaying fit, described in the text.}
\label{Ks_mass_down}
\end{figure}

\begin{figure}
\includegraphics[width=75mm]{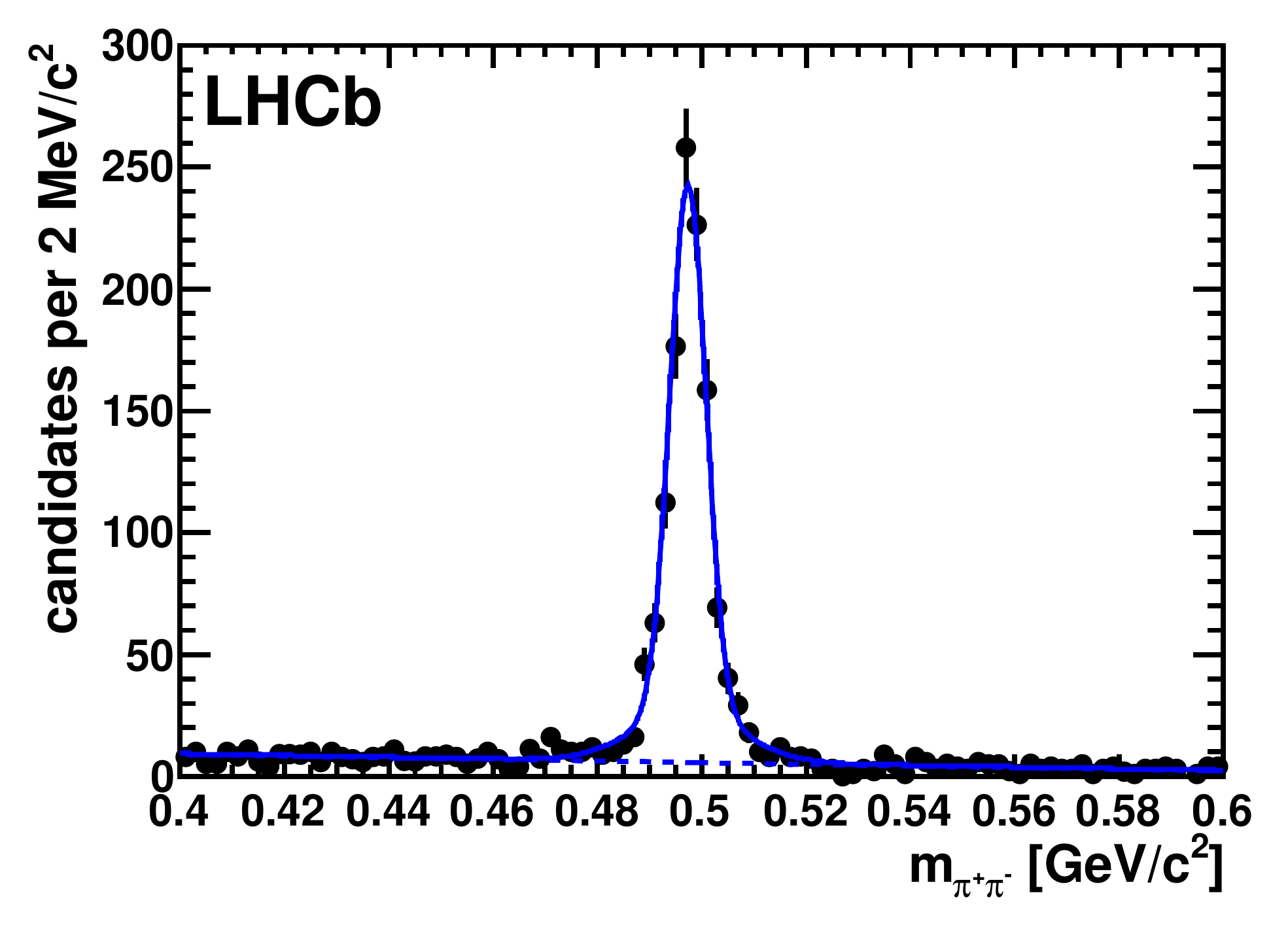}
\caption{Mass distribution of all selected $K_s$ candidates in the long-track analysis with the overlaying fit, described in the text.}
\label{Ks_mass_long}
\end{figure}

$K_{s}$ candidates were reconstructed from any combination of two oppositely charged tracks, assumed to be pions.
Due to the relatively long lifetime of the $K_{s}$ and the open VELO position, most of the tracks have no VELO information. Therefor the 
$K_{s}$ were selected using two different approaches: \\
The downstream-track selection uses tracks with hits in the tracking stations (TT and T1-T3) only, ignoring any VELO information. The selection is based on track quality 
cuts and requires the reconstructed $K_{s}$ to point back to the primary vertex, while the daughter tracks should have a large impact parameter.\\
The long-track selection uses tracks with hits in the VELO and the tracking stations (TT and T1-T3). The selection is based on the impact parameters 
of the  $K_{s}$ and their daughters using the reconstructed primary vertex. \\
The signal yield in each $(p_{\rm T}, y)$ bin was extracted using a $\chi^2$ fit to the invariant $m_{\pi \pi}$ of the two tracks mass with a linear function to describe the background and the sum of two gaussian 
functions to describe the signal yield. To account for $K_s$ produced from collisions with the beam and remaining gas, a beam-gas subtraction 
was made. More details are given in \cite{Kspaper}. 
The fitted mass distributions for the downstream-track and long-track analysis for the full $p_{\rm T}$ and $y$ range are shown in Fig. \ref{Ks_mass_down} and Fig.\ref{Ks_mass_long}.

\subsection{Results}
\begin{figure}
\includegraphics[width=75mm]{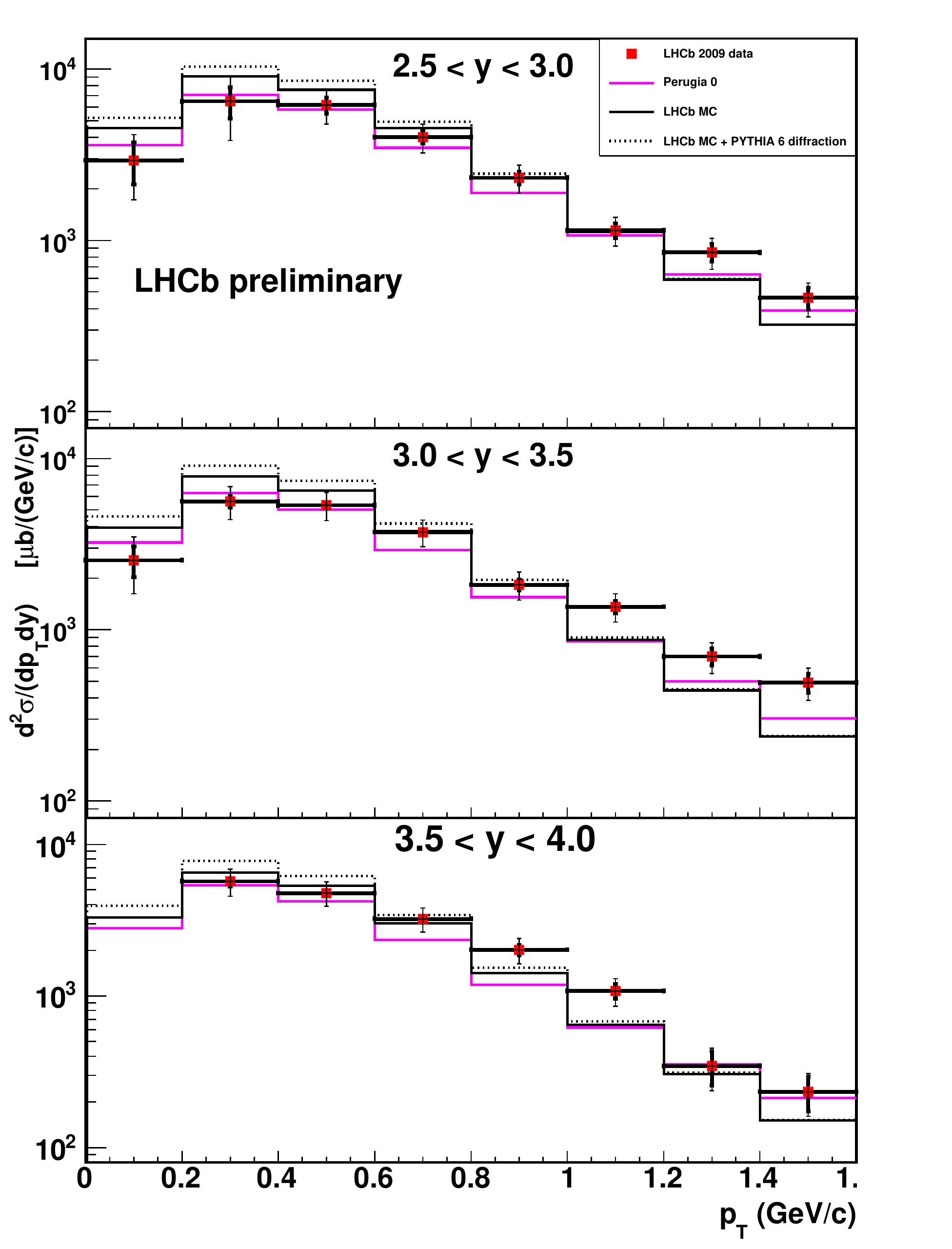}
\caption{Differential $K_s$ production cross section at $\sqrt{s} = 900\; \rm GeV$ as function of $p_{\rm T}$ in different bins of the rapidity. The thick 
error bars correspond to the statistical, the thin error bars to the total uncertainties. The histograms are different tunings of the PYTHIA generator 
(see text).}
\label{Ks_plot}
\end{figure}

The cross sections are calculated from both the downstream-track and the long-track analysis. Both lead to consistent results in each 
phase space bin. But as the two approaches are not statistically independent and the downstream-track analysis has the better statistical power 
the final results are taken from the downstream-track analysis (except the two lowest $p_{\rm T}$ bins) in Fig. \ref{Ks_mass_down}. The differential 
cross sections are shown in Fig. \ref{Ks_plot} as a function of the transverse momentum $p_{\rm T}$ for three different rapidity ranges. 
They are compared to Monte Carlo predictions using PYTHIA 6.4 with three different tunings \cite{pythia}:
\begin{itemize}
 \item the standard settings used by LHCb, including soft diffraction (black dashed line)
 \item the standard LHCb settings, excluding diffractive events (black solid line)
 \item the so-called Perugia 0 settings, not including any diffraction (pink line) \cite{perugia}
\end{itemize}

Dominating systematic uncertainties of the measurement are systematics on the determination of the luminosity ($\sim 15\%$), the agreement between data 
and Monte Carlo for the efficiency determination ($\sim 10\%$), the fit stability ($\sim 4\%$) and the stability of the selection cuts ($\sim 4\%$).

\section{$\frac{\bar{\Lambda}}{\Lambda}$ production ratio measurement}
A preliminary measurement of the $\frac{\bar{\Lambda}}{\Lambda}$ production ratio was performed with data recorded in 2010. 
The analysis was performed at the center of mass energies $\sqrt{s} = 900\; \rm GeV$ and $\sqrt{s} = 7\; \rm TeV$ using data samples with an integrated 
luminosity $L = 0.31\; \rm nb^{-1}$ and $L = 0.2\; \rm nb^{-1}$ respectively. 
All subdetectors were in operation. For the lower energy point the VELO halves were, as in 2009, retracted from their nominal position, this time by $10\; \rm mm$.\\
Data samples with two different magnetic field polarizations (upward and downward polarization) were combined for this measurement and compared with each other 
for systematic checks.

\subsection{Event selection}
The $\Lambda$ and $\bar{\Lambda}$ are reconstructed in the modes $\Lambda \rightarrow p\pi^-$ and 
$\bar{\Lambda} \rightarrow p^-\pi^+$. 
In the selection the $\Lambda$ candidates are identified by building the invariant mass of two oppositely charged track.
Only tracks with hits in the VELO and the tracking stations (TT and T1-T3) were used. 
The selection requires the $\Lambda$ or $\bar{\Lambda}$ to point back to the primary vertex to minimize the fraction of 
particles produced in material interactions. In addition cuts on the track quality and the impact parameter were applied.

\subsection{Results}
\begin{figure}
\includegraphics[width=75mm]{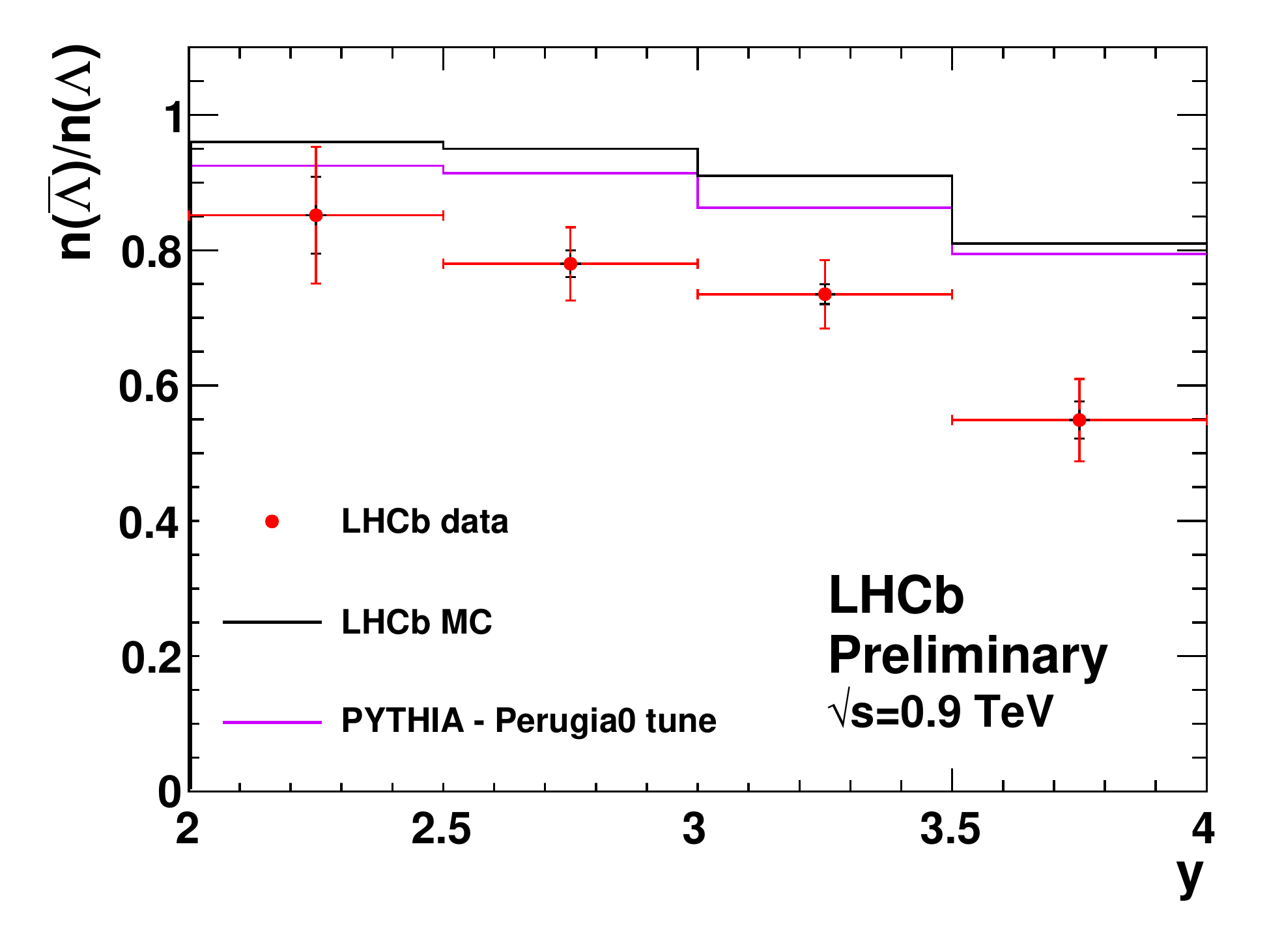}
\caption{$\frac{\bar{\Lambda}}{\Lambda}$ production cross section at a center of mass energy of $\sqrt{s} = 900\; \rm GeV$ 
as a function of the rapidity $y$. The error bars show the total uncertainty.The histograms show different Monte Carlo 
models (see text).}
\label{lambda_result_900}
\end{figure}

\begin{figure}
\includegraphics[width=75mm]{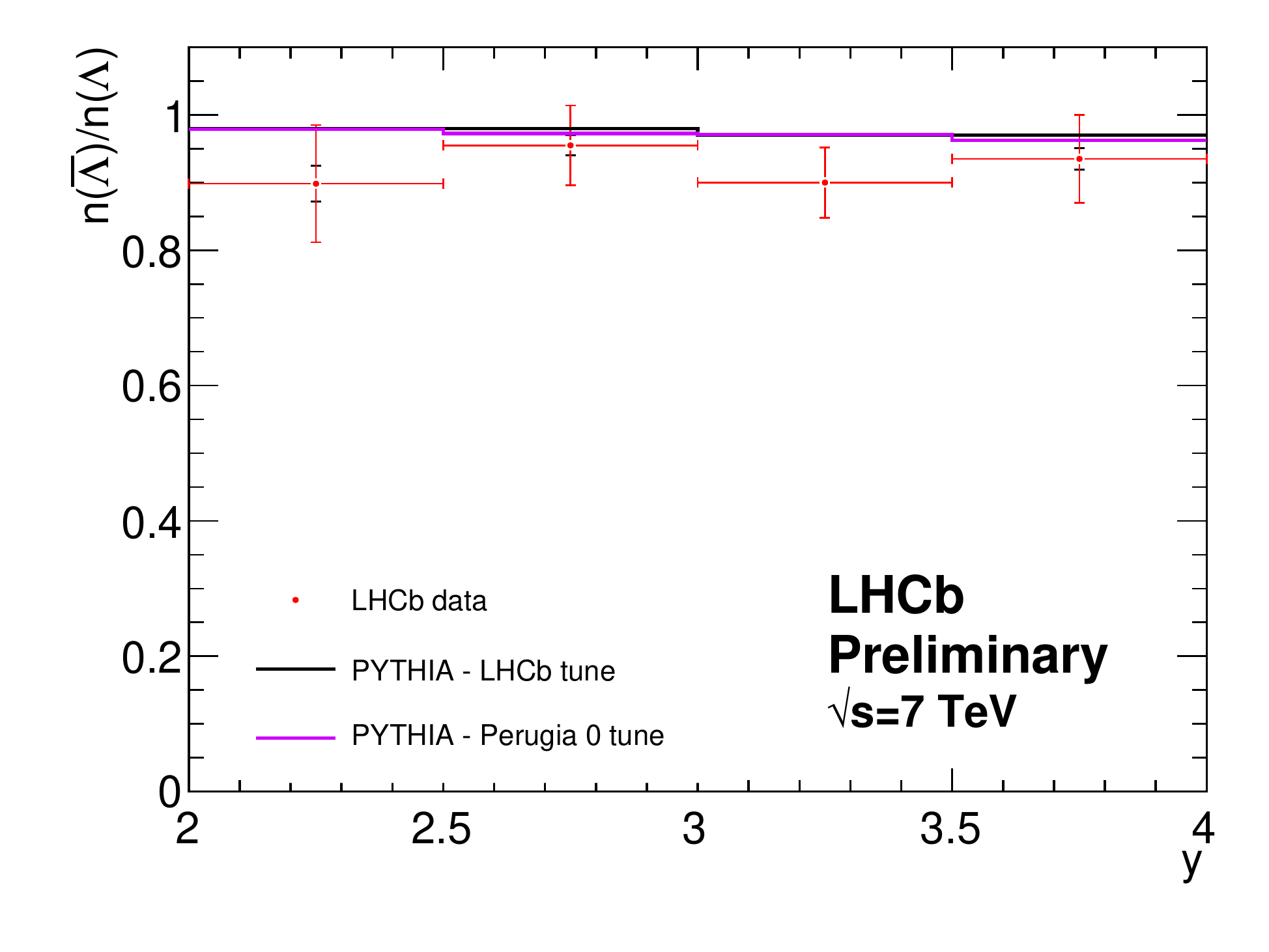}
\caption{$\frac{\bar{\Lambda}}{\Lambda}$ production ratio at a center of mass energy of $\sqrt{s} = 7\; \rm TeV$ 
as a function of the rapidity $y$. The error bars show the total uncertainty.The histograms show different Monte Carlo 
models (see text).}
\label{lambda_result_7}
\end{figure}

\begin{figure}
\includegraphics[width=75mm]{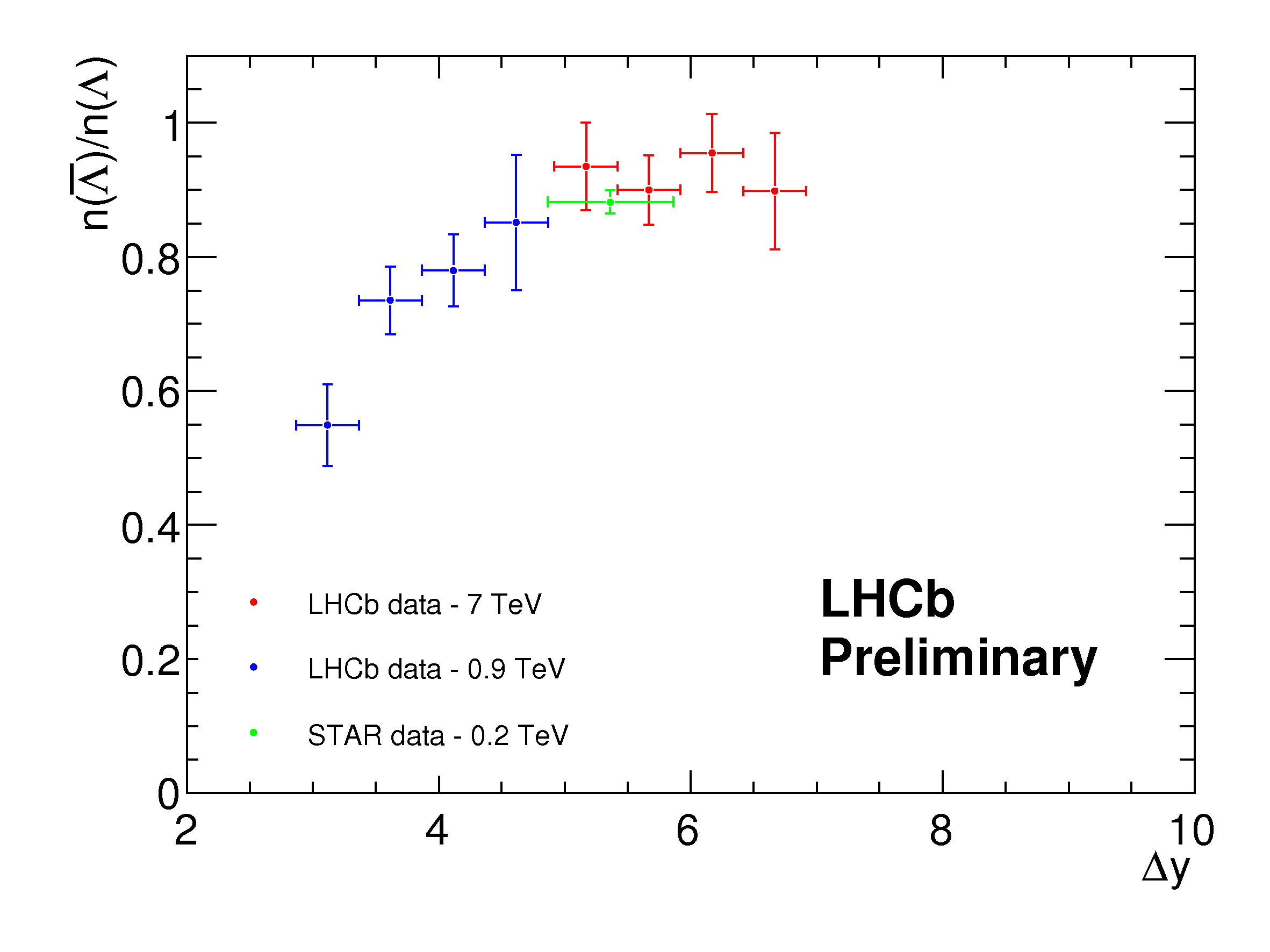}
\caption{$\frac{\bar{\Lambda}}{\Lambda}$ production ratio at center of mass energy of $\sqrt{s} = 7\; \rm TeV$ and $\sqrt{s} = 900\; \rm GeV$ 
as a function of the difference between beam and $\Lambda$ rapidity (see text).}
\label{lambda_comp}
\end{figure}

The results of the measurement are shown in Fig.\ref{lambda_result_900} for $\sqrt{s} = 900\; \rm GeV$ and 
in Fig. \ref{lambda_result_7} for $7\; \rm TeV$. They are compared with Monte Carlo predictions using two different PYTHIA 6.4 settings: the 
standard LHCb settings (black line) and the Perugia 0 settings (pink line). A preliminary examination of the systematic uncertainties 
leads to a $2\%$ relative and additional $0.02$ absolute uncertainty, where the dominant parts are uncertainties of matching the Monte Carlo $p_T$ 
distribution to data and the uncertainty of the material interaction cross section for particle and antiparticle.\\
Both the results for $\sqrt{s} = 900\; \rm GeV$ and $\sqrt{s} = 7\; \rm TeV$ are also displayed in Fig. \ref{lambda_comp} as a function of the difference 
of the beam rapidity and the rapidity of the reconstructed $\Lambda$: $\Delta y = y(\rm beam) - y(\Lambda)$. This allows a direct comparison 
of data taken at different energies. In addition a measurement of the STAR collaboration is shown (green). 

\section{Conclusion}
Using the first data collected in 2009 and 2010 the LHCb detector has been shown to perform well. All subdetectors already exhibit performances 
very close to their nominal values. Due to the unique angular coverage in forward direction LHCb has, in addition to its flavour-physics program, also a good opportunity to 
measure interesting forward minimum bias physics. Studies of a $\rm K_{s}$ differential production cross section measurement at a center of mass energy 
$\sqrt{s} = 900\; \rm GeV$ were presented. The results show reasonable consistency with expectations based on the PYTHIA Monte Carlo generator.
In addition a preliminary study of the $\frac{\bar{\Lambda}}{\Lambda}$ production ratio was performed. The results for $\sqrt{s} = 900\; \rm GeV$ show 
small differences from the generator predictions.\\
The integrated luminosity of $L = 1\;\rm fb^{-1}$ expected in 2011 will also allow LHCb to perform the planned flavour-physics measurements with 
the potential to discover New Physics effects.
%


\bigskip 

\begin{thebibliography}{9}   
\bibitem{detector}
The LHCb Collaboration, ``The LHCb Detector at LHC'', JINST 3 (2008) S08005.

\bibitem{Kspaper}
The LHCb Collaboration, ``Prompt $K^0_s$ production in $pp$ collisions at $\sqrt{s} = 900 \rm GeV$'', 
Physics Letters B 693 (2010) pp. 69-80

\bibitem{pythia}
T. Sj\" orstrand, S.Mrenna and P.Skands, ``PYTHIA 6.4 physics and manual'', JHEP 05 (2006) 026.

\bibitem{perugia}
P.Z.Skands, ``The Perugia tunes'', CERN-PH-TH-2010-113, arXiv:1005.3457v1 [hep-ph], May 2010
\end{thebibliography}

\end{document}